**Derivation of analytical expressions for optical gain coefficient in bulk semiconductors**

Elif Somuncu

*Department of Physics, Faculty of Arts and Sciences, Gaziosmanpaşa University, Tokat, Turkey*

*Email:elf_smnc@hotmail.com*

**Abstract**

An analytical model is proposed to determine the expressions of the semiconductor optical gain coefficient by using the special functions. These expressions differ from the standard numerical calculations approaches and they involve sums of the polylogarithm and incomplete gamma functions. One of the calculation results presented here reproduces formulae developed in the study. The new formulae obtained here provide a sufficient way to calculate the semiconductor optical gain on the total carrier density and temperature in the semiconductor lasers. The obtained analytical expression gives an accurate and precise evaluation of the semiconductor gain function. The performance of the proposed formula is confirmed by its implementation of the different values of the parameter. The sufficiency obtained results of this approach was confirmed by the other literature data.

## 1. Introduction

One of the important problems in physics is to investigate of the physical and chemical properties of the semiconductor laser. It is well known, the semiconductor laser , that occur of complex multi-layer structures requiring nanometer-scale precision and attentive design, play an important role in the electronic communications, especially a natural transmitter of digital data and interface for fiber optic cables [1-4]. Theoretical and experimental studies on semiconductor laser have been proposed for years by many researchers. Many effective and useful pieces of knowledge have been obtained from suggested theoretical and experimental studies. Suemune et.al. have indicated photopumped ZnSe/ZnSSe blue semiconductor and calculated theoretically optical gain [5]. Gavrielides et. al. have suggested analytically the stability boundaries of a semiconductor laser [6]. Mukai et.al. have investigated semiconductor laser diode with experimental [7]. Duarte et.al have a study on the monochromatic solutions of a semiconductor laser and its stability [8]. Zhang et.al have proposed a new method to drive semiconductor laser for optical frequency standard based on the constant voltage source [9]. Bao et.al have suggested WDM-based bidirectional chaotic communication for the semiconductor lasers system [10]. Mahmoud et.al have presented a theoretical work on the effect of intermodal asymmetric gain suppression on the dynamics and output of multimode semiconductor lasers [11]. Semiconductor laser theoretical definition is significant not only from a main point of view but also to generate new and advanced designs. The theoretical analysis and prediction of laser action based on a semiconductor diode was proposed first time in the study [12, 13]. As known, the notion of a semiconductor laser has been introduced by Basov et. al [12]. Basov et. al. have presented a work for quantum mechanical semiconductor [12]. Basov et. al. have a study on production of negative-temperature cases in P-N junctions of degenerate semiconductor [13]. With compared other lasers, the main advantage of semiconductor laser have a high optic gain and variation of the refractive index of the medium brings the change of the gain [14, 15]. Optical gain, which defines the optical amplification in the semiconductor material, is the most considerable necessity for the realization of a semiconductor laser. Therefore the efficient calculation method is important to calculate semiconductor gain on the total carrier density and temperature in the semiconductor lasers. There is only one study related to analytical evaluation of the semiconductor optical gain coefficient in the literature, to our knowledge [16]. As can be seen from work [16], the semiconductor optical gain coefficient evaluated full analytically with the help of binomial expansion theorem in case of $q>p$ and $s<0$. We note that numerical methods [17-21] have applied for approximate evaluation of the semiconductor optical gain coefficient [22-26]. It is well known that the numerical integration methods give correct results for a restricted range of parameters and are not convenient for most circuit designers since they are time consuming [27-32]. Therefore, the computational accuracy in the evaluation of the semiconductor optical gain coefficient is the very significant component of this kind of work.

This study presents a new analytical method for calculate the semiconductor optical gain coefficient for case arbitrary values of parameters. By using series expansion formulas the calculation of the semiconductor optical gain coefficient has been reduced to evaluation basic special functions. The obtained formulae seem to be easy to handle to make the numerical calculation and in this sense, their results would be useful to apply a realistic system.

## 2. Analytical expression for the semiconductor optical gain coefficient

The theoretical definition of semiconductor lasers properties has been investigated according to the Free-Carrier theory and the Fermi-Dirac distributions. By using the Free-Carrier theory and the Fermi-Dirac distributions the authors [33, 34] gives the following significant relation for the optical gain coefficient as following:

$$g = \frac{v|\mu_k|^2 \hbar\gamma}{4\pi^2\varepsilon_0 nc}\left(\frac{2m_r}{\hbar^2}\right)^{3/2} \int_0^\infty \frac{f_e(x)-f_v(x)}{(\hbar\gamma)^2+(x-\hbar\delta)^2}\sqrt{x}\,dx \qquad (1)$$

where $\hbar$ is Planck constant, $\gamma$ is homogeneous line width factor, $n$ is refractive index, $m_r$ is the reduced mass, and the $f_\alpha$ is the electrons and valance band Fermi-Dirac distribution defined as:

$$f_\alpha(x) = \frac{1}{\exp\left[\beta\left(\frac{m_r}{m_\alpha}x - \mu_\alpha\right)\right]+1} \qquad (2)$$

where $\mu_\alpha$ is the carrier quasi-chemical potential, $\beta = \frac{1}{kT}$ and $\alpha \equiv \upsilon$ and $e$ corresponds to electrons and valance band. The optical gain coefficient, considering Eq. (2) in Eq. (1) can be expressed as:

$$g = \frac{v|\mu_k|^2 \hbar\gamma}{4\pi^2\varepsilon_0 nc}\left(\frac{2m_r}{\hbar^2}\right)^{3/2}\left[e^{\beta\mu_e}Q(\hbar\gamma,\hbar\delta,\beta m_r/m_e,\beta\mu_e)+e^{\beta\mu_v}Q(\hbar\gamma,\hbar\delta,\beta m_r/m_h,\beta\mu_e)\right] \qquad (3)$$

where $Q$ is the semiconductor gain function can be written as

$$Q = (p,q,r,s) = \int_0^\infty \frac{\sqrt{x}}{\left[p^2+(x-q)^2\right](e^{rx}+e^s)}\,dx \qquad (4)$$

In the quantum statistical theory of semiconductor gain, it is a major significant the more correct assessment of semiconductor gain function because they are very susceptible to little errors. Also, it is important to note that the accurate assessment of the semiconductor optical gain coefficient is dependent to the evaluation of the Q functions. Therefore, the choice of reliable expressions for the semiconductor gain function is of major significance for accurate and precise calculations of the optical gain coefficient. To evaluate the semiconductor optical gain, we use in Eq. (4) the following exponential series expansion and binomial expansion theorems as [35]

$$e^{\pm x} = \sum_{k=0}^{\infty} (\pm 1)^k \frac{x^k}{k!}, \qquad (5)$$

$$(x \pm y)^n = \sum_{m=0}^{n} (\pm 1)^m F_m(n) x^{n-m} y^m \qquad (6)$$

where $F_m(n)$ is binomial coefficients described by

$$F_m(n) = \begin{cases} \dfrac{n(n-1)\ldots(n-m+1)}{m!} & \text{for integer } n \\ \dfrac{(-1)^m \Gamma(m-n)}{m!\Gamma(-n)} & \text{for noninterger } n \end{cases} \qquad (7)$$

The analytical expressions for the semiconductor optical gain coefficient have been derived by using the series expansion as following:

$$Q(p,q,r,s) = \lim_{N \to \infty} \sum_{i=0}^{N} \sum_{j=0}^{N'} (-1)^{i+j} F_j(i)(2q)^j \left(p^2+q^2\right)^{-(1+i)} T_{2i-j+1/2}(r,s) \qquad (8)$$

where $F_j(i)$ is binomial coefficients and $T_m(r,s)$ is the auxiliary function defined as:

$$T_m(r,s) = \int_0^{\infty} \frac{x^m}{e^{rx} + e^s} dx. \qquad (9)$$

The auxiliary functions $T_m(r,s)$ can be expressed as:

for $s \leq -1$

$$T_m(r,s) = -\frac{\Gamma(m+1) Li_{m+1}(-e^s)}{e^s r^{m+1}} \qquad (10)$$

for $s > -1$

$$T_m(r,s) = \frac{1}{r^{m+1}} \left[ \frac{s^{m+1}}{m+1} + \lim_{N \to \infty} \sum_{l=0}^{N} (-1)^{l+m} \frac{e^{-s(1+l)} \gamma(m+1,-ls)}{l^{m+1}} + \lim_{N' \to \infty} \sum_{k=0}^{N'} (-1)^k e^{ks} \frac{\Gamma(m+1,(k+1)s)}{(k+1)^{m+1}} \right] \qquad (11)$$

The special functions $Li_n(x)$, $\Gamma(\alpha,x)$ and $\gamma(\alpha,x)$ occurring in Eqs. (10) and (11) are the incomplete polylogarithm and incomplete gamma functions and defined as, respectively [35]:

$$Li_n(x) = \frac{1}{\Gamma(n)} \int_0^\infty \frac{t^{n-1}}{e^t/x - 1} dt \qquad (12)$$

$$\Gamma(\alpha,x) = \int_x^\infty t^{\alpha-1} e^{-t} dt \qquad (13)$$

$$\gamma(\alpha,x) = \int_0^x t^{\alpha-1} e^{-t} dt \qquad (14)$$

**Numerical Results and Discussion**

The semiconductor gain coefficient is the main important to evaluate theoretically for a semiconductor laser. Therefore, we have presented efficient methods for the analytical evaluation of integrals semiconductor gain function. An approximation for the analytical assessment of optic gain function has been derived and applied. The analytical formula is evaluated by the solution of the Eq. (4) that are calculated in readily and fast by using series expansion theorems. The method is based on exponential and the binomial expansion series formula that establishes an analytical solution in the form of a convergent series. The application of the semiconductor gain function is performed mostly using a numerical approach that calculation for important parameters and ensures the required results. In order to compare correctness and precision, the numerical integration, the available literature data [16], and the suggested analytical formula were applied in Mathematica Software 7.0. To show the precision and accuracy of the analytical expression, we offer several calculations of semiconductor gain function. The precision of the proposed method is admissible and can be propounded for the evaluation of the optical gain coefficient. By comparing the results of calculation with numerical integration and literature data [16], we indicated the validity of the given results. The obtained results are given Table 1-2. As seen in Table 1-2, the analytical expression is correct in the range of wide parameters. The results in Table 1-2 indicate that the method is extremely accurate and suitable for comparison purposes. As seen from Table 1-2, the results obtained from the semiconductor gain function are in very good agreement with numerical integration results and theoretical data [16]. As seen from Table 2, the results of calculation of the analytical expression are in very good agreement with the numerical integral calculation. Good agreement with analytical, theoretical, and numerical values has raised to the reliability of our obtained analytical formula. Table 3 indicates that the convergence properties of Eq. (8) are considered to change widely. As seen from Table 3, Eq. (8) shows the fastest convergence to the numerical results for various values of parameters. Also, in Table (3) displays the most rapid convergence to the numerical result, with seventeen digits stable and correct by the fiftieth terms in the infinite summation.

In consequence, a new analytical expression for the computation of the optical gain in semiconductor lasers has been presented in this work. The novelty of this work is that it is more efficient and performed accurately for the semiconductor gain function in the arbitrary parameter values. The newly obtained analytical formula for the semiconductor optical gain coefficient well prevents the calculation challenges.

**Conclusions**

In this work, we have been derived into explicit and efficiently analytical formula for the semiconductor gain function. The results from the analytical formula for arbitrary values of parameters are in good agreement with numerical results and literature data. These calculations are widely used in the semiconductor laser. In conclusion, in arbitrary values of parameters, the analytical formula offers the advantage of direct and precise calculation of the semiconductor gain function.

Table 1. Comparative calculation results of $Q(p,q,r,s)$ semiconductor gain function

| $p$ | $q$ | $r$ | $s$ | Eq.(8) | Ref.[16] | Mathematica Numerical Results |
|---|---|---|---|---|---|---|
| 5.2 | 2.6 | 4.5 | -0.2 | $2.3057694655302776672 \times 10^{-3}$ | $2.30575210705633379 \times 10^{-3}$ | $2.305752106867286450 \times 10^{-3}$ |
| 8.5 | 2.4 | 3.5 | -0.6 | $1.5097886900053914664 \times 10^{-3}$ | $1.5097886898999967 \times 10^{-3}$ | $1.509788689899996769 \times 10^{-3}$ |
| 12 | 11 | 10 | -0.6 | $9.10083926566079244 \times 10^{-5}$ | $9.10083926534021059 \times 10^{-5}$ | $9.1008392653402108 \times 10^{-5}$ |
| 18.5 | 12.4 | 13.5 | -0.8 | $3.1574477043853786 \times 10^{-5}$ | $3.157447704385142307 \times 10^{-5}$ | $3.1574477043851413 \times 10^{-5}$ |
| 23.8 | 15.2 | 15.6 | -0.9 | $1.5961348806248191 \times 10^{-5}$ | $1.596134880624276426 \times 10^{-5}$ | $1.5961348806242761 \times 10^{-5}$ |
| 25.2 | 22.6 | 21.5 | -0.95 | $6.8970202955913217 \times 10^{-6}$ | $6.897020295591316584 \times 10^{-6}$ | $6.8970202955591307273 \times 10^{6}$ |
| 31.8 | 22.6 | 14.5 | -0.4 | $8.717475292109295253384 \times 10^{-6}$ | $8.164666180471492413 \times 10^{-6}$ | $8.717475145766489230 \times 10^{6}$ |
| 43.1 | 32.2 | 24.1 | -0.45 | $2.1522973513670842630927 \times 10^{-6}$ | $2.15229734359733320475 \times 10^{-6}$ | $2.152297343597333768 \times 10^{-6}$ |

Table 2. Comparative calculation results of $Q(p,q,r,s)$ semiconductor gain function

| $p$ | $q$ | $r$ | $s$ | Eq.(8) | Mathematica Numerical Results |
|---|---|---|---|---|---|
| 6.2 | 3.6 | 4.3 | -0.2 | $1.62182136204011 \times 10^{-3}$ | $1.62180917327795 \times 10^{-3}$ |
| 7.2 | 2.6 | 5.3 | -0.5 | $1.0638314441811 \times 10^{-3}$ | $1.06383144337586 \times 10^{-3}$ |
| 12.5 | 11.5 | 10 | -0.55 | $8.29608538287100 \times 10^{-5}$ | $8.29608538163736 \times 10^{-5}$ |
| 17.2 | 15.3 | 13 | -0.8 | $3.13123019536 \times 10^{-5}$ | $3.13123019536565 \times 10^{-5}$ |
| 21.3 | 17.6 | 16 | -0.25 | $1.46370748786195 \times 10^{-5}$ | $1.46370501167247 \times 10^{-5}$ |
| 28.7 | 25.6 | 21 | -0.6 | $5.3004263127510 \times 10^{-6}$ | $5.30042631256278 \times 10^{-6}$ |
| 33.2 | 28 | 18 | -0.9 | $5.43779648898374 \times 10^{-6}$ | $5.43779648898101 \times 10^{-6}$ |
| 46.1 | 40.9 | 20.5 | -0.7 | $2.16703097270170891 \times 10^{-6}$ | $2.16703097269828 \times 10^{-6}$ |
| 50 | 45 | 42 | -0.75 | $6.2362583941772714 \times 10^{-7}$ | $6.23625839417505 \times 10^{-7}$ |
| 56 | 48 | 53 | -0.85 | $3.702850248778548 \times 10^{-7}$ | $3.70285024877852 \times 10^{-7}$ |
| 63.5 | 42.5 | 56 | -0.65 | $3.09772040337034 \times 10^{-7}$ | $3.0977204033940 \times 10^{-7}$ |

**Table 3.** Convergence of expressions Eq. (5) for $Q(p,q,r,s)$ as functions summation limits $N$ and $N'$

| $N = N'$ | p=56; q=48; r=53; s=-0.85     Eq.(8) |
|---|---|
| 5 | $3.7018187617749075 \times 10^{-7}$ |
| 10 | $3.702856658603566 \times 10^{-7}$ |
| 15 | $3.702850195087703 \times 10^{-7}$ |
| 20 | $3.7028502492958273 \times 10^{-7}$ |
| 25 | $3.70285024877308 \times 10^{-7}$ |
| 30 | $3.7028502487785485 \times 10^{-7}$ |
| 35 | $3.702850248778488 \times 10^{-7}$ |
| 40 | $3.7028502487784887 \times 10^{-7}$ |
| 45 | $3.7028502487784887 \times 10^{-7}$ |
| 50 | $3.7028502487784887 \times 10^{-7}$ |